\pdfoutput=1

\documentclass[12pt,letterpaper]{article}

\usepackage{jheppub}
\usepackage{verbatim}

\usepackage{graphicx}
\usepackage{subfigure}
\input{epsf}
\usepackage{epsfig}
\usepackage{epstopdf}
\usepackage{amsthm}

\newcommand{\labell}[1]{\label{#1}}
\def\({\left(} \def\){\right)}
\def\[{\left[} \def\]{\right]}
\def\al{\alpha} \def\bt{\beta}
\def\del{{\partial}}

\def\M{\mathcal{M}}

\newcommand{\non}{\nonumber \\}

\newcommand{\be}{\begin{equation}}
\newcommand{\ee}{\end{equation}}
\newcommand{\bea}{\begin{eqnarray}}
\newcommand{\eea}{\end{eqnarray}}
\newcommand{\ba}{\begin{eqnarray}}
\newcommand{\ea}{\end{eqnarray}}

\newcommand{\beq}{\begin{equation}}
\newcommand{\eeq}{\end{equation}}
\newcommand{\beqa}{\begin{eqnarray}}
\newcommand{\eeqa}{\end{eqnarray}}
\newcommand{\beqar}{\begin{eqnarray*}}
\newcommand{\eeqar}{\end{eqnarray*}}

\newcommand{\reef}[1]{(\ref{#1})}

\newcommand{\eg}{{\it e.g.,}\ }
\newcommand{\ie}{{\it i.e.,}\ }

\newcommand{\mt}[1]{\textrm{\tiny #1}}

\newcommand{\veps}{\varepsilon}

\newcommand{\A}{\mathcal{A}}

\newcommand{\C}{\mathcal{C}}

\newcommand{\cM}{\mathcal{M}}

\newcommand{\R}{\mathcal{R}}

\newcommand{\hr}{\hat\rho}
\newcommand{\hro}{\hat\rho_0}

\newcommand{\ha}{{a}}
\newcommand{\hb}{{ b}}
\newcommand{\hc}{{c}}
\newcommand{\hd}{{d}}
\newcommand{\he}{{e}}
\newcommand{\w}{\omega}

\title{Entanglement Entropy: A Perturbative Calculation}

\author{Vladimir Rosenhaus}
\emailAdd{vladr@berkeley.edu}
\affiliation{Center for Theoretical Physics and Department of Physics,\\
 University of California, Berkeley, CA 94720, U.S.A. }

\author{and Michael Smolkin}
\emailAdd{smolkinm@berkeley.edu}

\abstract{

We provide a framework for a perturbative evaluation of the reduced density matrix. The method is based on a path integral in the analytically continued spacetime. It suggests an alternative to the holographic and `standard' replica trick calculations of entanglement entropy. We implement this method within solvable field theory examples to evaluate leading order corrections induced by small perturbations in the geometry of the background and entangling surface. Our findings are in accord with Solodukhin's formula for the universal term of entanglement entropy for four dimensional CFTs.

}

\begin{document}
\maketitle

\section{Introduction}

Entanglement entropy is a rapidly developing technique in condensed matter physics \cite{Levin:2006zz,cardy0} and holography \cite{rt1,rt2}. One of the main theoretical gaps that substantially limits  its studies is  the paucity of computational tools. In this paper we construct a perturbative framework for computing entanglement entropy of the vacuum purely within the context of quantum field theory (QFT). 

As of today the existing tools for computing entanglement entropy include: the replica trick, conifolds, and the elegant prescription of Ryu and Takayanagi \cite{rt1,rt2}. The replica trick, and its generalizations, is the only generic approach to calculating entanglement entropy within field theory \cite{cardy0,wilczek}. It rests on evaluating the partition function on an $n$-folded cover of the background geometry where a cut is introduced throughout the exterior of the entangling surface. However, evaluation of the partition function on a replicated manifold can only be carried out in a limited number of cases \cite{Hertzberg:2010uv}. On the other hand, the Ryu-Takayanagi prescription is much easier to implement. It plays a central role in characterizing new properties of holographic field theories, \eg \cite{kleb}, and provides new insights into the quantum structure of spacetime \cite{van,bian,Myers:2013lva}.  Recently,  the generalized replica trick 
was successfully implemented in the bulk AdS space to provide strong evidence for the Ryu-Takayanagi conjecture \cite{Lewkowycz:2013nqa}.\footnote{For precursors, see also \cite{furry} and critique of \cite{furry} in \cite{head}.}

In \cite{CHM} Casini, Huerta and Myers showed that the reduced density matrix for spherical entangling surfaces in flat space is conformally equivalent to a thermal state on the hyperbolic geometry, and that the entanglement entropy equals  the thermodynamic entropy of this thermal state. This observation provided an alternative derivation of the holographic entanglement entropy for spherical regions in flat space. However, their construction tightly relies on the conformal symmetry of the boundary CFT and on the (spherical) geometry of the entangling surface. Hence, their work raises a natural question: how does one accommodate small disturbances of their framework within a perturbative approach? In this paper we propose a Euclidean path integral formalism that addresses this question. In particular, our method paves the way for an alternative approach to calculating entanglement entropy within quantum field theory. 

In Section \ref{sec2} we set aside holography, the replica trick, and other existing methods of calculating entanglement entropy and begin with the `standard' Euclidean path integral definition of the reduced density matrix. Next, we foliate spacetime in the vicinity of the entangling surface in such a way as to encode both the geometric structure of the surface and the geometry of the background. This choice of coordinates is one of the central aspects of our approach,  as any deformation can be now thought of as a background deformation. As a result, a perturbative framework around systems with known reduced density matrices is established. We finish this section with analysis of small perturbations induced by relevant deformations of the QFT.

In Section \ref{rindler} we consider the entanglement entropy obtained by dividing the field theory into two (semi-infinite) regions with a single flat plane separating them. In this case the entanglement entropy  for  any QFT equals the thermal entropy observed by an accelerating Rindler observer  \cite{kabastra}. We apply our general formalism to calculate leading order corrections induced by either slight curvature of the background or mild deformations of the flat wall separating the two subsystems. In particular, we evaluate the universal divergence of the entanglement entropy induced by these modifications in four dimensional spacetimes. The results are in complete agreement with the structure of the universal terms in entanglement entropy of 4D conformal field theories originally proposed by Solodukhin \cite{solo}.\footnote{See also \cite{scone,Dong:2013qoa} for a recent derivation based on the squashed cones technique and \cite{Solodukhin:1994yz} for early studies of the logarithmic divergences in the context of black hole entropy.} 

The main focus of Section \ref{sec4} is the analysis of perturbations around spherical entangling surfaces. The unperturbed case in the context of QFT was studied in \cite{CHM}, whereas in this work we implement our formalism to investigate consequences of small perturbations. The resulting corrections to the universal divergence of entanglement entropy in 4D match known results in the literature \cite{solo}.

\section{General framework} 
\label{sec2}

We start with a general quantum field theory that lives on a  $d$-dimensional Euclidean manifold $\M$ equipped with a Riemannian metric $g_{\mu\nu}$. The action of the field theory is given by $I_0(\phi, g_{\mu\nu})$, where $\phi$ collectively denotes all the QFT fields. We assume that the system resides in the vacuum state\footnote{For entanglement entropy of excited states in the holographic context see, \cite{Hubeny:2007xt,Blanco:2013joa}, whereas the path integral approach to this problem is elaborated on in \cite{klich}. }. The entangling surface is chosen to be some general $(d-2)$-dimensional surface $\Sigma$. Our notation for the rest of the paper is summarized in Appendix \ref{not}. 

\begin{figure}[tbp] 
\centering
%\subfigure[]{
	\includegraphics[width=2in]{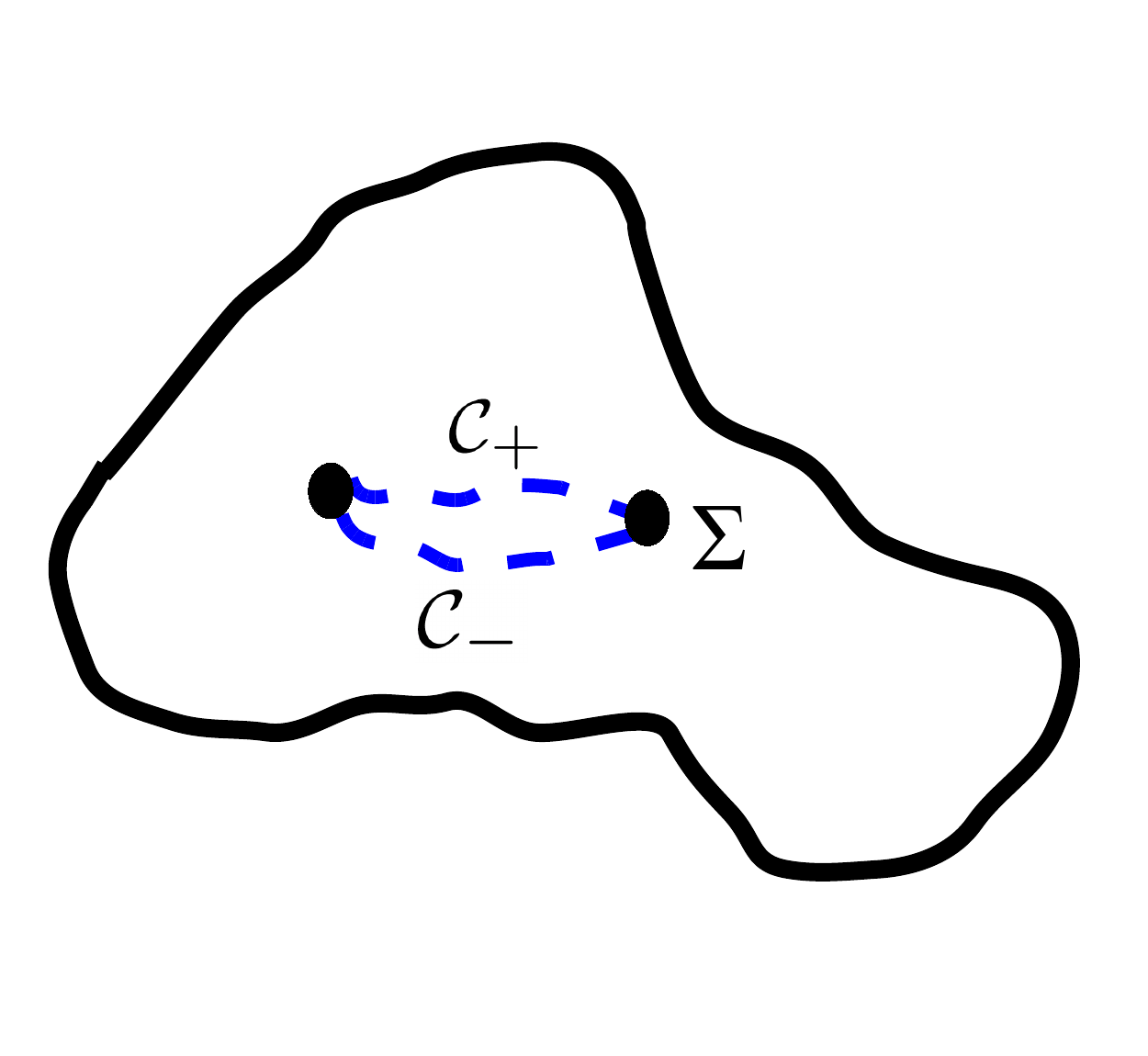}
%	}
\caption{Abstract sketch of the two dimensional transverse space to the entangling surface $\Sigma$.  $\mathcal{C}_\pm$ are the two sides of the cut $\mathcal{C}$ where the values $\phi_\pm$ of the field $\phi$ are imposed.}
\label{figGen}
\end{figure}

The degree of entanglement between the QFT degrees of freedom inside and outside of $\Sigma$ is encoded in the reduced density matrix $\rho_0$ that can be written as a path integral over $\M$ with a $(d-1)$-dimensional cut $\mathcal{C}$, such that $\del\,\mathcal{C}=\Sigma$ 
\be
 [\rho_0]_{\phi_-\phi_+}\equiv \langle\phi_{-} | \rho_0 | \phi_+ \rangle=\int_{\phi(\C_+)=\phi_+  \above 0pt \phi(\C_-)=\phi_- } \mathcal{D}\phi \, e^{-I_0(\phi, g_{\mu\nu})}~,
 \labell{reden}
\ee
where $\C_\pm$ are the two sides of the cut and $\phi_{\pm}$ are some fixed field configurations (see Fig.~\ref{figGen}). 

In general, evaluation of the above path integral is not a tractable problem, but there are exceptions, \eg planar and spherical surfaces in $R^{d}$ that we are going to explore later. For the rest of this section the details of $\rho_0$ are not crucial, we only need to assume that it is known, since the main purpose is to get a closed form expression for small perturbations of $\rho_0$ as a consequence of slight deformations of the background metric $g_{\mu\nu}$ and  entangling surface $\Sigma$, or perturbations of the QFT by, \eg a relevant operator.

We start with the normalized density matrix,
\be
\hat\rho_0={\rho_0 \over \text{Tr} \rho_0}~.
\ee  
The corresponding modular Hamiltonian, $\hat K_0$, and the entanglement entropy,  $S_0$, are given by
\bea
 \hat K_0&=&-\log\hat\rho_0 ~,
 \non
 S_0&=&-\text{Tr}\hat\rho_0\log\rho_0 ~.
\eea
Now let us consider a perturbation of $\hat\rho_0$ by a small amount $\delta\hat\rho$,
\be
 \hat\rho=\hat\rho_0+\delta\hat\rho~,
\ee
The new density matrix $\hat \rho$ is assumed to be normalized, and therefore $\text{Tr}\, \delta\hr = 0$. The corresponding modular Hamiltonian, $\hat K$, and the entanglement entropy,  $S$, can be constructed perturbatively provided that $\hro$ and $\delta\hr$ are known
\be 
 S=-\text{Tr}\hat\rho\log\hr=S_0+\text{Tr}(\delta\hr \, \hat K_0) - {1\over 2} \text{Tr}(\delta\hr\,\hro^{-1}\,\delta\hr)  + \mathcal{O}(\delta\hr^3) ~.
 \labell{modEE}
\ee
We note that the expression \reef{modEE} should, via the Baker-Cambell-Hausdorff formula, include terms involving commutators. We have, however, suppressed such terms as for our applications these terms are contact terms, and in cases where it is relevant it will be implicitly assumed that one accounts for contact terms appearing in correlation functions.
%We should note that in the expression (\ref{modEE}) one should, via the  Baker-Campbell-Hausdorff formula, also include terms involving the commutator of $\delta\hr$ %and $\hat K_0$.\footnote{The proper treatment of such terms will be explored in a later publication. However, these commuators are not as threatening as they may %appear. For instance, for a planar $\Sigma$ in $\mathbb{R}^d$, the modular Hamiltonian is local and the relevant commutators are between local operators and %therefore only have support at coincident points. For this case, it was shown in \cite{Rosenhaus:2014ula} that for relevant deformations of a QFT, the second order %correction to the entanglement entropy agrees with \reef{modEE} (where it is implicitly assumed that one accounts for the various contact terms that may emerge if the %insertion points of operators within the expectation value coincide).} However, to first order in $\delta\hr$ such terms are irrelevant \cite{Blanco:2013joa}.
To first order in $\delta\hr$ the above expansion reveals a `first law' of entanglement entropy \cite{Bhattacharya:2012mi,Blanco:2013joa,klich}
\be
 \delta S=\text{Tr}(\delta\hr \, \hat K_0)=\delta\langle K_0 \rangle~.
 \labell{1law}
\ee

In those examples that we are going to consider, it is possible (but not always necessary) to implement a conformal transformation that maps the background $\M$, and hence the path integral \reef{reden}, onto $S^1\times H^{d-1}$ which we will denote as $\mathcal{H}$. Of course, we implicitly restrict our consideration here to CFTs. Remarkably, under this transformation the entangling surface $\Sigma$ is mapped onto the conformal boundary of $H^{d-1}$ while fixed states $|\phi_\pm\rangle$ are mapped onto constant slices $\tau_\mt{E}=0$ and $\tau_\mt{E}=\beta$ (see Section \ref{sec4} and Fig.~\ref{fig:Hyp} there). The latter condition ensures that under this map the reduced density matrix $\hro$ transforms into a normalized thermal density matrix $\hr_T$ on $\mathcal{H}$. In particular, $S^1$ plays the role of Euclidean time, $\tau_\mt{E}$, and its period is identified with the inverse temperature $\beta$. Additionally,
\be
 \hr_T= \hat U \, \hro \, \hat U^{-1}~,
 \labell{dentran}
\ee
where $\hat U$ is a unitary CFT operator that implements the conformal transformation. For example, the primary spinless operators, $\mathcal{\hat O}$,  of the CFT locally transform as\footnote{The subscript on $\mathcal{\hat O}$ indicates on which manifold the operator has support.}
\be
 \mathcal{\hat O}_{\mathcal{H}}=\Omega^\Delta \, \hat U \, \mathcal{\hat O}_{\M} \, \hat U^{-1}~,
\ee
where $\Delta$ is the scaling dimension of $\mathcal{\hat O}$ and $\Omega$ is the conformal factor that relates the metrics on the two manifolds
\be
 ds^2_{\M}=\Omega^2 ds^2_{\mathcal{H}}~.
 \labell{confmap}
\ee
In what follows we consider separately perturbations of the QFT action, and perturbations associated with either slight changes in the background geometry or mild deformations of the entangling surface $\Sigma$.

\subsection{Geometric perturbations}

In general, the modular Hamiltonian depends on the background geometry as well as on the geometry of the entangling surface. The same is true about conformal transformations of $\M$ onto $\mathcal{H}$ that relate the density matrices as in   \reef{dentran}. Such mappings are sensitive to changes in the background geometry as well as to deformations of the entangling surface $\Sigma$. While the former sensitivity is obvious, the latter follows from the fact that   \reef{dentran} is valid if and only if the field configurations $\phi_{+}$ and $\phi_-$ are mapped onto constant slices $\tau_\mt{E}=0$ and $\tau_\mt{E}=\beta$, respectively. Therefore the mapping, if it exists, certainly depends on the details of $\Sigma$.
 
These observations lead us to construct a special foliation of $\M$ that encodes both the background geometry as well as the structure of the entangling surface \cite{Lewkowycz:2013nqa, Dong:2013qoa}. Such a foliation for a generic $\M$ and $\Sigma$ can only be found perturbatively in the distance from the entangling surface. Sufficiently far from $\Sigma$ caustics may be encountered and our coordinate system will break down. However, this region is not relevant for us.  We present here the final answer for the foliation, with the details  relegated to Appendix \ref{fol}. To second order in the distance from $\Sigma$, the metric on $\M$ is given by
\bea
 d s^2_{\M}&=& (\delta_{\ha\hb}-{1\over 3}\R_{\ha\hc\hb\hd}|_\Sigma x^{\hc}x^{\hd})dx^\ha dx^\hb 
 + \big(A_i+{1\over 3}x^\hb \veps^{\hd\he}\R_{i\hb\hd\he}\big|_{\Sigma} \big)\veps_{\ha\hc}\, x^\ha dx^\hc dy^i
 \non
 &+& \Big(\gamma_{ij}+2 K_{\ha ij} \, x^\ha+  x^\ha x^{\hc}\big( \delta_{\ha\hc} A_iA_j+\R_{i\ha \hc j}|_\Sigma +K_{\hc \, i l} K_{\ha\,j}^{~l}  \big) \Big)dy^idy^j + \mathcal{O}(x^3)~,
 \non
 \labell{ans}
\eea
where $\{y^i\}_{i=1}^{d-2}$ and $\{x^\ha\}_{\ha=1}^{2}$ parametrize $\Sigma$ and the 2-dimensional transverse space, respectively. The entangling surface $\Sigma$ is located at $x^\ha=0$ and $\gamma_{ij}$ is the corresponding induced metric, $\veps_{\ha\hc}$ is the volume form of the transverse space, whereas $\R_{\mu\nu\al\bt}$ and $K^\ha_{ij}$ are the background and extrinsic curvatures, respectively. Finally, $A_i$ is the analog of the Kaluza-Klein vector field associated with dimensional reduction over the transverse space. Note that by construction the structure of $\Sigma$ is built into the above ansatz.

The ansatz for the metric with a slightly perturbed background and mildly modified entangling surface $\Sigma$ can be obtained by varying   \reef{ans} around the unperturbed background. In particular, the metric will take the following form
\be
  g_{\mu\nu}=\bar g_{\mu\nu} + h_{\mu\nu}~,
  \labell{met}
\ee
where $\bar g_{\mu\nu}$ is the unperturbed metric of the form \reef{ans} with known coefficients, while $h_{\mu\nu}$ contains all the information about perturbations that occurred in the background and entangling surface geometries.  

If $\Sigma$ is everywhere a small deformation of the original entangling surface, \eg if it is a plane everywhere except that in some localized region there is a small ``bump'', then perturbative analysis applies globally on $\Sigma$. However, $h_{\mu \nu}$ does not necessarily even need to be small everywhere on the entangling surface. If, for example, the surface does not globally look like a plane by having a low curvature but long turn, then we can implement a cut and paste procedure suggested in~\cite{bian}. We cut the surface along regions which are sufficiently flat, compute the entanglement entropy for each section, and then paste the results together. Of course, this cut and paste procedure is not straightforward and there are potential computational subtleties that need to be addressed.

Substituting decomposition (\ref{met}) into the path integral representation of the density matrix, (\ref{reden}), and expanding the result around  $\bar g_{\mu\nu}$ yields,
\bea
 [\hr]_{\phi_-\phi_+}&=&{1\over \mathcal{N}}\int_{\phi(\C_+)=\phi_+  \above 0pt \phi(\C_-)=\phi_- } \mathcal{D}\phi \, e^{-I_0(\phi, \bar g_{\mu\nu} + h_{\mu\nu})}
 \non
 &=& {1\over \mathcal{N}} 
 \int_{\phi(\C_+)=\phi_+  \above 0pt \phi(\C_-)=\phi_- } \mathcal{D}\phi \, e^{-I_0(\phi, \bar g_{\mu\nu} )} \big(1+{1\over 2}\int_{\M} T^{\mu\nu}_\M h_{\mu\nu} + \ldots \big) ~,
   \labell{redden2}
\eea
where $T^{\mu\nu}_\M$ is the energy-momentum tensor of the QFT on the unperturbed Euclidean manifold $\M$
\bea
 T^{\mu\nu}_\M&=&-{2\over \sqrt{\bar g}} {\delta I_0 \over \delta \bar g_{\mu\nu}}~.
 \labell{emt}
\eea
The normalization constant $\mathcal{N}$ appearing in (\ref{redden2}) is given by
\begin{multline}
 \mathcal{N}= 
 \int\mathcal{D}\phi_0\int_{\phi(\C_+)= \phi(\C_-)=\phi_0 } \mathcal{D}\phi \, e^{-I_0(\phi, \bar g_{\mu\nu} )} \big(1+{1\over 2}\int_{\M} T^{\mu\nu}_\M h_{\mu\nu} + \ldots \big) 
 \\
 =\mathcal{N}_0\(1+{1\over 2}\int_{\M} \langle \hat T^{\mu\nu}_\M \rangle_0 h_{\mu\nu} + \ldots\) ~,
   \labell{N}
\end{multline}
where $\langle \hat T^{\mu\nu}_\M\rangle_0$ is the expectation value of the stress tensor in the state $\hro$, while $\mathcal{N}_0$ is the normalization constant of the unperturbed density matrix $\hro$, 
\be
 \mathcal{N}_0=\int\mathcal{D}\phi_0\int_{\phi(\C_+)= \phi(\C_-)=\phi_0 } \mathcal{D}\phi \,  \, e^{-I_0(\phi, \bar g_{\mu\nu} )}~.
    \labell{N0}
\ee

 It is convenient to think of the path integral in \reef{redden2}, \reef{N} and \reef{N0} as an effective evolution from the slice $\C_+$ to the slice $\C_-$ \cite{klich}. In particular, based on these equations one can write 
\begin{multline}
 [\delta\hr]_{\phi_- \phi_+}  =\langle\phi_{-} | \delta\hr | \phi_+ \rangle=
 {1\over 2} \int_{\M} \langle \phi_-, \theta_f | \hat{\mathcal{U}}(\theta_f,\theta) \, \hat T^{\mu\nu}_\M(\theta)\, \hat{\mathcal{U}}(\theta,\theta_i)|\phi_+, \theta_i\rangle h_{\mu\nu} 
 \\
 -  {1\over 2}  [\hr_0]_{\phi_- \phi_+}  \int_{\M}\langle \hat T^{\mu\nu}_\M\rangle_0 h_{\mu\nu}(\theta) ~,
 \labell{varho}
\end{multline}
where we have used the definition $\delta\hr=\hr-\hro$, $\theta$ is the polar angle around the entangling surface such that $\theta_i$ and $\theta_f$ correspond to  the slices $\C_+$ and $\C_-$ respectively, and $\hat{\mathcal{U}}$ is the evolution operator. In general, $\hat{\mathcal{U}}$ has a complicated structure. If, however, the unperturbed background is such that the undeformed entangling surface exhibits rotational symmetry in the transverse space, then this symmetry will be inherent in the path integral representation of  $\hro$. In particular, as shown in  \cite{kabastra} (see also \cite{rindler,Unruh, mark}) in this case $\hat K_0$ is identical to the generator of angular evolution around $\Sigma$ and $\hat{\mathcal{U}}$ takes the form 
\be
 \hat{\mathcal{U}} (\theta_2,\theta_1) = \exp\big( -{\theta_2-\theta_1\over 2\pi} \, \hat K_0 \big) ~.
\ee
Stripping off the field states in \reef{varho}, yields
\be
 \delta\hr  =
 {1\over 2} \int_{\M} \hat{\mathcal{U}}(\theta_f,\theta) \, \Big(\hat T^{\mu\nu}_\M(\theta) - \langle \hat T^{\mu\nu}_\M\rangle_0\Big) \hat{\mathcal{U}}(\theta,\theta_i) h_{\mu\nu}  ~.
 \labell{varho1}
\ee
The entanglement entropy across $\Sigma$ now reads
\be
 S=S_0+{1 \over 2}\int_{\M} \langle \hat T^{\mu\nu}_\M \hat K_0 \rangle_c \, h_{\mu\nu} + \ldots~,
 \labell{varEE0}
\ee
where $\langle \ldots \rangle_c$ is the connected two point function in the state $\hro$. We should note that our result (\ref{varEE0}) is valid for a general field theory, and is not necessarily restricted to a CFT.

Moreover, if we restrict our consideration to conformal field theories, then it is possible to generalize the above results to include the case when the state undergoes a conformal mapping as in \reef{dentran}, \reef{confmap}. We first recall the rule for conformal transformation of the energy-momentum tensor, 
\be
 T^{\mu\nu}_{\M}=\Omega^{-d-2}{\del x^\mu \over \del X^{\alpha}}{\del x^\nu \over \del X^\bt} \big( T^{\al\bt}_{\mathcal{H}} + \A^{\al\bt}  \big) ~,
 \labell{anomEMT}
\ee
where $X^\mu$ are coordinates on $\mathcal{H}$,  $x^\mu$ collectively denotes $(x^\ha, y^i)$ and $\A^{\mu\nu}$ is the higher dimensional analog of the Schwarzian derivative. Hence, from   \reef{redden2} we obtain 
\be
 [\hat U \hr \, \hat U^{-1}]_{\tilde\phi_+ \tilde\phi_-}= {1\over \mathcal{N}} 
 \int_{\phi(\tau_\mt{E}=0)=\tilde\phi_+  \above 0pt \phi(\tau_\mt{E}=\bt)=\tilde\phi_- } \mathcal{D}\phi \, e^{-I_0(\phi, \bar g_{\mu\nu} )}
  \bigg(1+{1\over 2}\int_{\mathcal{H}} \Omega^{-2}\big(T^{\mu\nu}_\mathcal{H}+\A^{\mu\nu}\big) h_{\mu\nu} + \ldots \bigg) ~,
  \labell{redden3}
\ee
where $\tilde\phi_\pm$ are the conformally transformed field configurations $\phi_\pm$, 
\be
 | \tilde\phi_\pm\rangle = \hat U | \phi_\pm\rangle ~.
\ee
Also note that the normalization constant $\mathcal{N}$ in    \reef{N} can be rewritten as
\be
 \mathcal{N}=  \mathcal{N}_0\(1+{1\over 2}\int_{\mathcal{H}} \Omega^{-2}\langle \hat T^{\mu\nu}_\mathcal{H}\rangle_T h_{\mu\nu} 
 +{1\over 2}\int_{\mathcal{H}} \Omega^{-2}\A^{\mu\nu} h_{\mu\nu}+ \ldots\) ~,
 \labell{NT}
\ee
where $\langle \hat T^{\mu\nu}_\mathcal{H}\rangle_T$ is the thermal expectation value of the stress tensor on $\mathcal{H}$.
Combining eqs. \reef{redden3} and \reef{NT}, yields
\be
\hat U \delta\hr \, \hat U^{-1}={1\over 2}\, \int_{\mathcal{H}} \hat{\mathcal{U}}_T(\bt , \tau_{\mt E}) \Big(\hat T^{\mu\nu}_\mathcal{H}(\tau_{\mt E})
 - \langle \hat T^{\mu\nu}_\mathcal{H}\rangle_T\Big) \hat{\mathcal{U}}_T(\tau_{\mt E},0)\,\Omega^{-2} h_{\mu\nu} \, ,
 \labell{varho2}
\ee
where we used the transformation rule \reef{dentran}, and $\hat{\mathcal{U}}_T$ is the evolution operator on $\mathcal{H}$,
\be
 \hat{\mathcal{U}}_T (\tilde\tau_{\mt E},\tau_{\mt E}) = \exp\big( -(\tilde\tau_{\mt E} - \tau_{\mt E} ) \, \hat H \big) ~,
\ee
where $\hat H$ is the Hamiltonian that generates $\tau_\mt{E}$ translations. It is related to the modular Hamiltonian on $\M$ by $\hat K_0=\hat U^{-1}\bt\hat H \hat U$.

Since the von Newman entropy is invariant under unitary transformations, the entanglement entropy across $\Sigma$ can be evaluated using the density matrix on $\mathcal{H}$. Substituting \reef{varho2} into \reef{modEE}, yields
\be
 S=S_T+{\bt \over 2}\int_{\mathcal{H}} \Omega^{-2}\langle \hat T^{\mu\nu}_\mathcal{H}\hat H \rangle_c \, h_{\mu\nu} + \ldots~,
 \labell{varEE}
\ee
where $S_T$ is the thermal entropy of the CFT in the state $\hr_T$, while $\langle \ldots \rangle_c$ is the (thermal) connected two point function on $\mathcal{H}$. This result is simply a conformal transformation \reef{confmap} of \reef{varEE0}, accompanied by the rule \reef{anomEMT}.

\subsection{Relevant perturbations}

The main goal of this  subsection is to investigate the consequences of small perturbations of the QFT by, \eg relevant operators. The general form of the reduced density matrix \reef{redden2} that undergoes such a perturbation is  
\bea
 [\hr]_{\phi_+\phi_-}&=&{1\over \mathcal{N}}\int_{\phi(\C_+)=\phi_+  \above 0pt \phi(\C_-)=\phi_- } \mathcal{D}\phi \, e^{-I_0(\phi, \bar g_{\mu\nu})+g\int_\M \mathcal{O}}
 \non
 &=& {1\over \mathcal{N}} 
 \int_{\phi(\C_+)=\phi_+  \above 0pt \phi(\C_-)=\phi_- } \mathcal{D}\phi \, e^{-I_0(\phi, \bar g_{\mu\nu} )} \(1+g\int_{\M} \mathcal{O} 
 + {g^2\over 2} \big(\int_{\M} \mathcal{O}\big)^2 + \ldots \) ~,
\eea
where $g$ is the coupling constant, the scaling dimension of $\mathcal{\hat O}$ is $\Delta<d$, and we assume that the effect of the deformation is small, \eg the theory sits sufficiently close to the UV fixed point.

The normalization constant this time is given by
\be
 \mathcal{N}=  \mathcal{N}_0\(1+ g\int_{\M} \langle\mathcal{O} \rangle_0
 +{g^2\over 2}\int_\M\int_\M \langle \mathcal{\hat O}\mathcal{\hat O}\rangle_0 + \ldots\) ~,
\ee
where the expectation values are taken in the vacuum state. Following now the same steps as in the previous subsection, we obtain the leading order correction to $S_0$,
\be
 \delta S= g \int_\M \langle \mathcal{\hat O} \hat K_0\rangle_c ~.
\ee

If the unperturbed theory is a CFT and the entangling surface is either a plane or a sphere, then the leading correction to $S_0$ vanishes since $\hat K_0 \sim \hat T_{\mu\nu}$ and therefore $\langle \hat K_0 \mathcal{\hat O} \rangle_c=0$. Hence, in this case we have to resort to the second order perturbation.  Using \reef{modEE} yields\footnote{It was verified in \cite{Rosenhaus:2014ula} that the second order terms in \reef{modEE} are legitimate in the case of a plane in flat space. For a general entangling surface the modular Hamiltonian is expected to be nonlocal, and there may be subtleties with the appropriate treatment of contact terms and with the use of \reef{relcorr}.},
\be
 \delta S={g^2\over 2}\int_\M\int_\M \Big( \langle \hat K_0 \mathcal{\hat O}\mathcal{\hat O}\rangle_c -  \langle \mathcal{\hat O}\mathcal{\hat O}\rangle_c\Big)~.
 \label{relcorr}
\ee
%where certain caution should be taken before evaluating the expectation value of the two point function in the above expression, since according to \reef{modEE} this %correlator is evaluated using a path integral with an effective interval of evolution that has to be three times bigger than the interval of evolution used to compute the %three point function in \reef{relcorr}. 

We finish this section with a comment that it would be interesting to compare the results based on \reef{relcorr} with the holographic predictions made in \cite{Hung:2011ta} and with the field theory computations in \cite{largeN} where the deformations of critical points by relevant operators were studied. We hope to report on this in a forthcoming publication.

\section{Perturbations of a planar entangling surface} 
\labell{rindler}

\begin{figure}[tbp] 
\centering
%\subfigure[]{
	\includegraphics[width=2in]{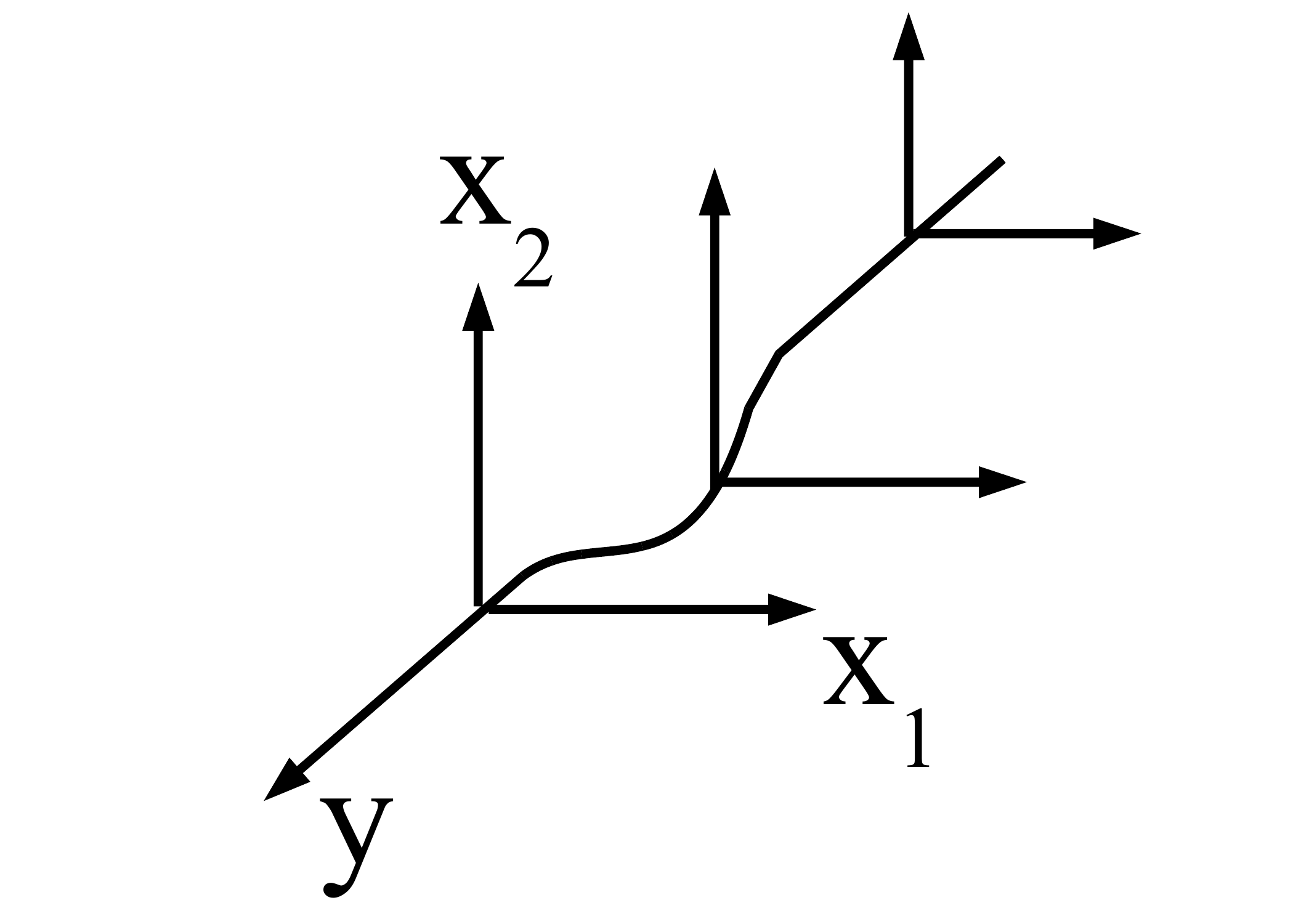}
%	}
\caption{A sketch of a slightly deformed entangling surface (curved line) in three dimensions. ($x_1,x_2$) span the transverse space to $\Sigma$ , while $y$ parametrizes $\Sigma$. The foliation (\ref{ans}) is designed to capture the geometry of the neighborhood of a given entangling surface $\Sigma$.} 
\label{fig:3d}
\end{figure}

In this section we explore the leading order correction (\ref{varEE0}) in the case of small perturbations of a planar entangling surface in flat space. These perturbations could arise from the entangling surface being slightly deformed (see Fig.~\ref{fig:3d}), or if the background geometry is weakly curved. For simplicity we restrict our discussion to four spacetime dimensions and evaluate the logarithmic divergence of entanglement entropy. This divergence is universal since it is independent of the regularization scheme. 

The entanglement entropy of the unperturbed plane in flat space is closely related to the Unruh effect observed by a uniformly accelerating observer in Minkowski space. Indeed, the reduced density matrix for the vacuum for the semi-infinite domain $x_1>0$  is obtained by tracing out the region $x_1<0$ on a constant zero Minkowski time slice. This is precisely the region hidden by Rindler horizon and the resulting reduced density matrix has a thermal interpretation in the sense of Unruh \cite{Unruh,kabastra} with a space dependent temperature that scales as $x_1^{-1}$. A Rindler observer who is confined to the right wedge, and who is passing through $x_1$ at $t=0$, finds himself immersed in a thermal bath of Unruh radiation. The sum of thermal entropies observed by all Rindler observers is the entanglement entropy, and the divergence of the temperature as $x_1\rightarrow 0$ gives rise to the UV divergence of entanglement entropy. 

\begin{figure}[tbp] 
\centering
%\subfigure[]{
	\includegraphics[width=2in]{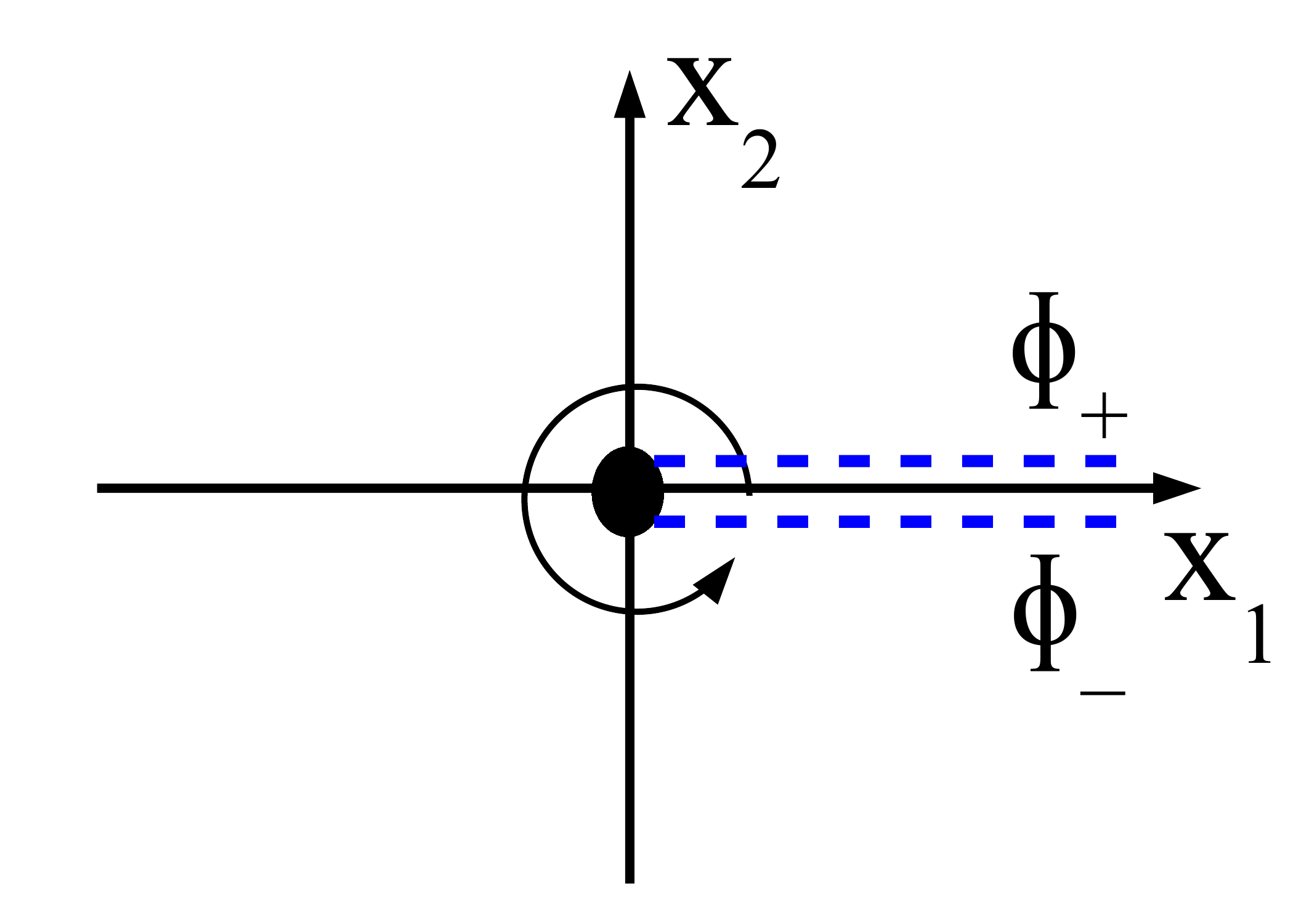}
%	}
\caption{ Transverse space to the entangling surface in the analytically continued spacetime.  $\Sigma$ is located at the origin. The reduced density matrix is given by a path integral \reef{reden} with fixed boundary conditions $\phi_+$ ($\phi_-$) on the upper (lower) dashed blue lines. }
 \label{fig:2d}
\end{figure}

Analytic continuation of the Rindler wedge to Euclidean signature maps it onto the entire Euclidean space  with a puncture at the origin. In Minkowski signature, this puncture corresponds to the Rindler horizon. Furthermore,  the analytically continued Rindler Hamiltonian, $\hat H_R$ , becomes the generator of rotations in the transverse space to $\Sigma$, and as shown in \cite{kabastra} the path integral (\ref{reden}) can be written as
\be
[\rho_0]_{\phi_+\phi_-}  = \langle\phi_{-} | e^{-2 \pi \hat H_R} | \phi_+ \rangle ~.
\labell{eq:Kab}
\ee 
In particular, we immediately deduce that the modular Hamiltonian is proportional to the Rindler Hamiltonian, $\hat K_0  = 2 \pi \hat H_R$, which plays the role of the angular evolution operator in the transverse space to $\Sigma$. (see Fig.~\ref{fig:2d})

What we have said so far is the standard story for flat space. In a general spacetime, since any region locally looks flat, we expect the leading divergence of the entanglment entropy will be insensitive to the background, in so much as that it scales as an area. The subleading terms of the entanglement entropy are dominated by the region near the entangling surface but have sensitivity to regions slightly away from it as well. 

Far away from the surface corrections to the background metric induced by perturbations of the system may be large. However, the further away some region is from the surface, the less relevant it is for the entanglement entropy. Stated in the language of accelerated observers: those who are highly accelerated and close to the Rindler horizon are unlikely to notice a large deviation from a thermal spectrum, while those with small acceleration who are far away find little Unruh radiation and the thermal effect is practically zero.

\subsection{Calculation}
The leading order correction to the entanglement entropy of a flat plane is given by (\ref{varEE0}),
\begin{equation}
\delta S = \pi\int_{R^4}{\langle T^{\mu \nu} H_R\rangle h_{\mu \nu}} .
 \labell{eq:deltaS}
\end{equation}
Here $H_R$ is the Rindler Hamiltonian in the unperturbed spacetime\footnote{The minus sign appears due to the definition \reef{emt} of the energy-momentum tensor in Euclidean signature.},
\begin{equation}
H_R = - \int_A {T_{\mu \nu} \xi^{\mu} n^{\nu} } ~,
\end{equation} 
where $A= \{x \in R^4 \, \big| x_2=0, x_1>0\}$, $\xi=x_1\del_2-x_2\del_1$ is the Killing vector field associated with rotational symmetry around the plane at $x^a=0$, while $n = \partial_2$ is normal to $A$. Thus, 
\begin{equation}
H_R = -\int_{A}{ \, x_1\, T_{22}} ~.
\end{equation}
Substituting $H_R$ into (\ref{eq:deltaS}) gives
\begin{equation} \label{eq:deltaS2}
\delta S = -\pi \int{d^2x\ d^2y\  d^2 \bar{y}\ d\bar{x}_1\ \bar{x}_1\ h^{\mu \nu}(x,y)\ \langle T_{\mu \nu}(x, y)T_{22}(\bar{x}, \bar{y})\rangle}~.
\end{equation}
Here the coordinates are $x^{\mu} = (x^{a},y^{i})$ where $x^{a}$ with $a=1,2$ are orthogonal to the entangling surface (see Fig.~\ref{fig:2d}) and $y^{i}$ with $i=1,2$ are along the entangling surface. Also, $\bar{x}_2=0$.
From (\ref{ans}) we find that there are two terms in $h_{\mu \nu}$ that are responsible for the logarithmically divergent contribution to $\delta S$. 
They are 
\begin{eqnarray}
h_{ij} &=& x^{a}x^{c}\mathcal{R}_{i a c j} \\
h_{ab} &=& -\frac{1}{3} \mathcal{R}_{a c b d}x^c x^d ~.
\end{eqnarray}
Note that the $\delta \gamma_{ij}$ term in (\ref{ans}) is not relevant as it contributes to the `area law' correction. Also, the cross terms $d x d y$ will give vanishing contributions. Finally,  terms proportional to the extrinsic curvatures contribute at second order within our perturbative expansion (since the extrinsic curvature of the plane is zero and the contribution of the linear term vanishes identically). 

The connected 2-pt function for the stress tensor for a CFT is given in \cite{Osborn:1993cr}, 
\begin{equation}
\langle T_{\mu \nu}(x,y) T_{2 2}(\bar{x},\bar{y})\rangle = \frac{C_T \, \mathcal{I}_{\,\mu \nu, 2 2}}{\left((x-\bar{x})^2 + (y-\bar{y})^2\right)^4}
\end{equation}
where 
\be
\mathcal{I}_{\mu \nu, 22} = I_{\mu 2}I_{\nu 2} - \frac{\delta_{\mu \nu}}{4} ~,
\label{eq:OsI}
\ee
with
\be
I_{\mu 2} = \delta_{\mu 2} - \frac{2 (x-\bar{x})_{\mu}\ x_2}{(x-\bar{x})^2 + (y-\bar{y})^2}~.
\ee
In Appendix \ref{sec:appC} we preform the integral (\ref{eq:deltaS2}) and find
\begin{equation} 
\delta S = \frac{c}{6 \pi}  \int{d^2y \left(\delta^{ac}\delta^{bd}\R_{abcd} + \delta^{ij} \delta^{ac} \mathcal{R}_{i a c j} \right)   \log(\ell/\delta)}~.
\label{eq:RindAnswer}
\end{equation}
Here $\ell$ is the characteristic scale of the perturbations, $\delta$ is the UV cut-off, and $C_T = 40 c/\pi^4$ with $c$ being the central charge of the CFT defined by the trace anomaly,
\be
\langle T^\mu_{~\mu} \rangle =\frac{c}{16\pi^2}\,\int_{\mathcal{M}} C_{\mu\nu\rho\sigma} C^{\mu\nu\rho\sigma} -\frac{a}{16\pi^2}\int_{\mathcal{M}}E_4\,, 
\labell{trace4}
 \ee
where $C_{\mu\nu\rho\sigma} $ is the Weyl tensor and $E_4$ is the Euler density, 
 \be
  E_4= R_{\mu\nu\rho\sigma} R^{\mu\nu\rho\sigma}-4R_{\mu\nu} R^{\mu\nu} + R^{2}~.
 \ee

Our correction \reef{eq:RindAnswer} should be compared with Solodukhin's formula \cite{solo} for the universal part of entanglement entropy in the case of a four dimensional CFT,
\be
 S_{CFT}={1\over 2\pi} \int_{\Sigma} \[ c \, ( \delta^{\ha\hc}\delta^{\hb\hd}C_{\ha\hb\hc\hd}+K^\ha_{ij} K^{ ij}_\ha-{1\over 2}K^\ha K_{\ha})-a \, \R^\Sigma \] \log(\ell/\delta) ~,
 \labell{4dEE}
\ee
where $\R^\Sigma$ is the intrinsic curvature of the entangling surface. Of course, for the case of a planar surface in flat space $S_{CFT}$ vanishes identically. 

Varying \reef{4dEE} around the flat plane embedded in $R^d$, we obtain to linear order in small perturbations
\begin{multline}
 \delta S_{CFT}={c\over 2\pi} \int_{\Sigma}  \,  \delta^{\ha\hc}\delta^{\hb\hd}C_{\ha\hb\hc\hd}  \, \log(\ell/\delta) 
 \\
 ={c\over 6\pi}   \int_{\Sigma} \( \delta^{ac}\delta^{bd}\R_{abcd}
 + \gamma^{ij}\delta^{\ha\hc} \R_{i\hat \ha\hc j} +  \gamma^{ij}\gamma^{kl} \R_{ikjl}\) \, \log(\ell/\delta)  ~,
 \labell{4dcorr2}
\end{multline}
where in the second equality we used the definition of the Weyl tensor. This expression matches   \reef{eq:RindAnswer} since the last term is a total derivative in this case, and therefore its integral vanishes. Indeed, the first variation of the Gauss-Codazzi relation \reef{gc} around the flat plane embedded in flat space gives
\be
 \gamma^{ij}\gamma^{kl} \R_{ikjl}|_\Sigma= \del^i(\del^j\delta\gamma_{ij} - \gamma^{mn}\del_i \, \delta\gamma_{mn})~,
 \labell{Rsig}
\ee 
where we have used the general variational rule
\be
 \delta\R_\Sigma=-\R^{ij}_\Sigma \, \delta\gamma_{ij} +  \nabla^i(\nabla^j\delta\gamma_{ij} - \gamma^{mn}\nabla_i \, \delta\gamma_{mn})~,
 \label{varule}
\ee
where $\nabla_i$ is covariant derivative compatible with the unperturbed induced metric $\gamma_{ij}$.

Before closing this section let us make a couple of comments. First, we note that \reef{eq:RindAnswer} and (\ref{4dcorr2}) are independent of the central charge $a$. This is a straightforward consequence of the fact that $R^\Sigma$ is the Euler density of a two-dimensional manifold, and therefore the last term in  \reef{4dEE} is a topological invariant that does not change under smooth deformations of the entangling surface and background, \ie
\be
\delta\int_\Sigma \R^\Sigma=\int_{\Sigma} \({1\over 2}\gamma^{ij}\R^\Sigma-\R^{ij}_{\Sigma}\)\delta\gamma_{ij}=0~,
\ee
where by assumption the deformed and original setups approach each other at infinity and we used the fact that $\Sigma$ is a two-dimensional manifold. 

Second, it should be noticed that terms in \reef{4dEE} that are quadratic in extrinsic curvature do not contribute to the leading order correction to the entanglement entropy since $K^a_{ij}$ of a flat plane vanishes. To see the effect of extrinsic curvatures one has to study second order perturbations within our formalism and this will be addressed in a forthcoming publication. In order to see the effect of extrinsic curvatures at first order, we now turn to spherical entangling surfaces.

\section{Perturbations of a spherical entangling surface} 
\label{sec4}

\begin{figure}[tbp] 
\centering
%\subfigure[]{
	\includegraphics[width=5in]{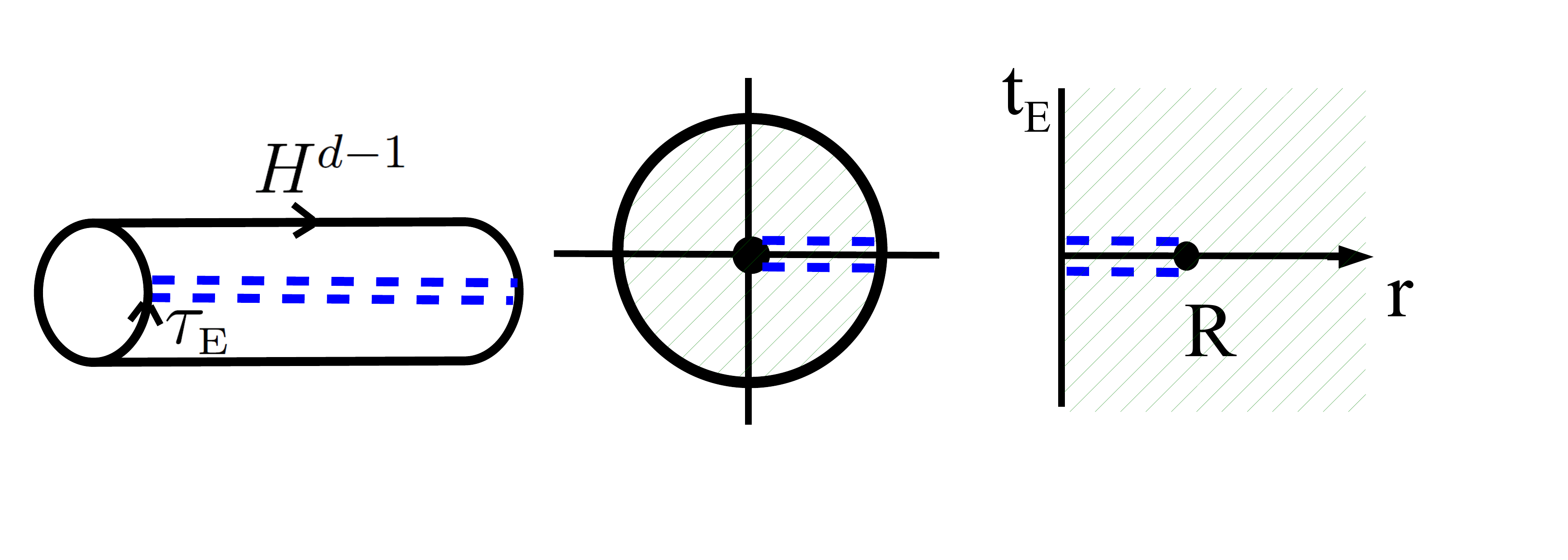}
%	}
\caption{We conformally transform between $\mathcal{H}$ (left) and $R^{d}$ (right). We first map from the $\sigma \equiv u +i \tau$ coordinates of $\mathcal{H}$ to $e^{-\sigma}$ (middle); here the origin is $u = \infty$ and the boundary circle is $u =0$. We then map via (\ref{magic}) to $R^{d}$. Dashed lines on the left represent $\tau_{\mt E}=0^+,\bt^-$ slices of $\mathcal{H}$ that are mapped through an intermediate step onto $t=0^\pm$ sides of the cut throughout the interior of the sphere $r=R$ on the right} 
\label{fig:Hyp}
\end{figure}

In this section the background manifold $\M$ will be identified with $R^d$, and the entangling surface $\Sigma$ will be a sphere, $S^{d-2}$, of radius $R$. We first show that there is a conformal map that transforms between Euclidean path integral representations of $\hro$ and $\hr_T$ and then apply the analysis of Sec.~\ref{sec2} to compute the first order corrections to the entanglement entropy due to slight deformations of $R^d$ and $S^{d-2}$.  

Let us recall that the partition function on $\mathcal{H}\equiv S^1 \times H^{d-1}$ can be evaluated by a path integral on
the Euclidean background 
 \be
ds^2_{\mathcal{H}}=d\tau_\mt{E}^2+ R^2\,\left(du^2+\sinh^2\!
u\,d\Omega^2_{d-2}\right)\,,
 \labell{hyper2}
 \ee
where the Euclidean time coordinate has period $\Delta \tau_\mt{E} =\bt = 2\pi R$. In the following, it will be convenient to introduce complex coordinates:
 \be
\sigma=u+i \tau_\mt{E}/R \qquad {\rm and}\qquad \w = r+i t_\mt{E}\, ,
 \labell{ccord}
 \ee
where the latter will be used below to describe a conformally mapped geometry. Note that both $u$ and $r$ are radial coordinates, and we must have
Re$(\sigma)=u>0$ and Re$(\w)=r>0$. With the first of these new
coordinates, the above metric \reef{hyper2} can be written as
 \be
ds^2_{\mathcal{H}}=R^2\( d\sigma\,d\bar{\sigma}+
\sinh^2\!\(\frac{\sigma+\bar{\sigma}}{2}\)\,d\Omega^2_{d-2}\)\,.
 \labell{hyper2a}
 \ee
Now we make the coordinate transformation \cite{renyi} (see Fig.~\ref{fig:Hyp})
 \be
 e^{-\sigma}=\frac{R-\w}{R+\w}\,.
 \labell{magic}
 \ee
Since we are considering $d\ge3$ there is no guarantee that this
holomorphic change of coordinates will result in a conformal
transformation. However, one can readily verify the above metric
\reef{hyper2a} becomes
 \bea
ds^2_{\mathcal{H}}&=&\Omega^{-2}\, \left[ d\w\,d\bar{\w}+ \(\frac{\w+\bar{\w}}{2}\)^2
d\Omega^2_{d-2}\right]
 \nonumber\\
&=&\Omega^{-2}\, \left[\, dt_\mt{E}^2 + dr^2+
r^2\,d\Omega^2_{d-2}\,\right]~,
 \labell{flat}
 \eea
where
 \be
 \Omega^{-1}=\frac{2R^2}{|R^2-\w^2|}=\cosh\,u + \cos ( \tau_\mt{E} / R)\,.
 \labell{confact}
 \ee
\begin{figure}[tbp] 
\centering
%\subfigure[]{
	\includegraphics[width=2.5in]{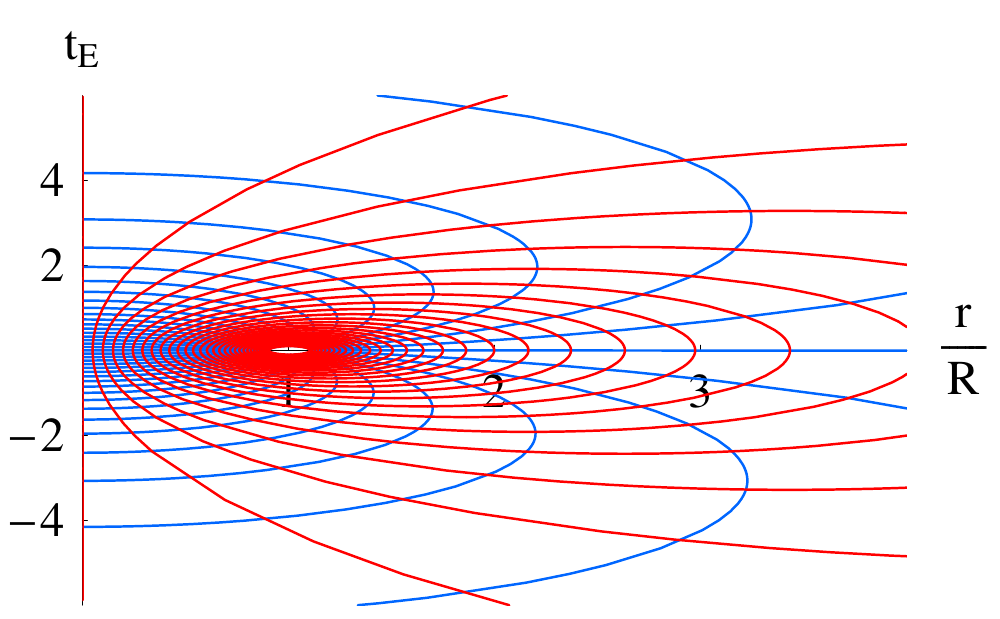}
%	}
\caption{We show the constant $\tau_{\mt E}$ slices (blue)  and constant $u$ slices (red) in the $(r,t_{\mt E})$ plane (\ref{magic2}). The sphere is located at $r/R=1$, $t_{\mt E}=0$ and corresponds to $u\rightarrow \infty$. The vertical line ($r=0$) corresponds to $u=0$.} \label{fig:plot}
\end{figure}
Hence, after eliminating the conformal factor $\Omega^{-2}$ in the second
line of  \reef{flat}, we recognize that the final line element is
simply the metric on $d$-dimensional flat space. Written explicitly in terms of real coordinates, \reef{magic} takes the form (see Fig.~\ref{fig:plot})
\begin{eqnarray}
r &=& R \, \frac{\sinh u}{\cosh u + \cos (\tau_\mt{E}/R)} \, , 
\non
t_\mt{E} &=& R \, \frac{\sin (\tau_\mt{E}/R)}{\cosh u+ \cos (\tau_\mt{E}/R)}~.
\labell{magic2}
\end{eqnarray}
Note that  \reef{magic2} can be obtained by analytic continuation to Euclidean time of the conformal mapping between causal domain of a sphere in Minkowski space and Lorentzian $\mathcal{H}$ \cite{CHM}. Under this analytic continuation the boundary of the causal domain shrinks to a sphere of radius $R$ while its interior spans the rest of Euclidean space. Note also that the conformal factor \reef{confact} is everywhere regular on the Euclidean space excluding the sphere of radius $R$.

Eq.~\reef{magic2} implements a simple bijection between $\mathcal{H}$ and $R^d$. Furthermore, the conformal boundary of the hyperbolic space $H^{d-1}$ is mapped onto a $(d-2)$-dimensional sphere of radius $R$ sitting on a $t_\mt{E}=0$ slice of $R^d$. Finally, constant time slices $\tau_\mt{E}=0^+$ and $\tau_\mt{E}=\bt^-$ are mapped respectively onto $t_\mt{E}=0^-$ and $t_\mt{E}=0^+$ of the cut $\C=\{x^\mu\in R^d \,| \, 0\leq r < R\,, ~ t_\mt{E}=0 \}$. Hence we have shown that the conformal map \reef{magic} transforms between the thermal state on $\mathcal{H}$ and the entangled state $\hro$ for a spherical region in $R^d$. 

In particular, the Hamiltonian on $\mathcal{H}$ is simply related to the modular Hamiltonian on $R^d$,
\be
 \hat K_0=\hat U^{-1}\bt\hat H \hat U=\bt\int_{H^{d-1}} \hat U^{-1} \, T^{\tau_{\mt E}\tau_{\mt E}}_{\mathcal{H}} \, \hat U ~.
 \label{modsph}
\ee
This expression agrees with the result of \cite{CHM}. Indeed, using eq.~\reef{magic}, we obtain
\be
\frac{\partial u}{\partial t_\mt{E}}=-{1\over R}\frac{\partial \tau_\mt{E}}{\partial r}={iR(\omega^2-\bar\omega^2)\over (R^2-\omega^2)(R^2-\bar\omega^2)}~ , ~
\frac{\partial u}{\partial r}={1\over R}\frac{\partial \tau_{\mt E}}{\partial t_\mt{E}}= {2R^3-R(\omega^2+\bar\omega^2)\over (R^2-\omega^2)(R^2-\bar\omega^2)}~,
\labell{CReq}
\ee
where the first equalities in the above expressions reveal the standard Cauchy-Riemann conditions. Now choosing for simplicity the slice $\tau_{\mt E}=0$ in \reef{modsph}, and using transformation rule \reef{anomEMT}, yields 
\be
 \hat K_0=2\pi \int \, {R^2-r^2\over 2 R}T^{t_{\mt E}t_{\mt E}}_{\M}  + c'~,
\ee
where the integral runs over the interior of the sphere of radius $R$, and $c'$ is some constant that ensures that the density matrix has unit trace.

\subsection{Geometric perturbations}

The metric on $\M$ is given by 
\be
 ds^2_{\M}=dt_\mt{E}^2 + dr^2+ r^2\,d\Omega^2_{d-2}~.
\ee
We rewrite it as
\be
 ds^2_{\M}=dx_1^2 + dx_2^2+ \(1+{2\over R} \, x_2 + {x_2^2\over R^2} \) \,ds^2_\Sigma~,
 \labell{Sans}
\ee
where we defined a new set of coordinates $t_\mt{E}=x_1\, , ~ r=R+x_2$ with  $-R\leq x_2 <\infty$, and $ds^2_\Sigma$ is the line element on a sphere of radius $R$
\be
 ds^2_\Sigma=R^2 d\Omega^2_{d-2}~.
\ee
The extrinsic curvatures of $\Sigma$ in this case are given by
\be
 K^{\hat 1}_{ij}=0\, , ~ K^{\hat 2}_{ij}={\gamma_{ij}\over R} ~,
\ee 
where $\gamma_{ij}$ is the induced metric on a sphere of radius $R$. 

We assume that the background curvature, induced metric, and extrinsic curvatures acquire corrections $\R_{\mu\nu\al\bt}\, , ~\delta \gamma_{ij}$ and $ \delta K^\hc_{ij}$ parametrized by some infinitesimal parameter $\epsilon$
\be
 R^2 \, \R_{\mu\nu\al\bt} \sim R \, \delta K^\hc_{ij} \sim \delta \gamma_{ij} \sim \epsilon
 \labell{varparam}
\ee
As a result, the slightly perturbed metric can be expressed in the form of \reef{met}, where $\bar g_{\mu\nu}$ is given by   \reef{Sans}, while $h_{\mu\nu}$ takes the form 
\bea
 h_{\mu\nu}dx^\mu dx^\nu&=&-{1\over 3}\R_{\ha\hc\hb\hd}|_\Sigma x^{\hc}x^{\hd}dx^\ha dx^\hb 
 + \big(A_i+{1\over 3}x^\hb \veps^{\hd\he}\R_{i\hb\hd\he}\big|_{\Sigma} \big) \veps_{\ha\hc}\, x^\ha dx^\hc dy^i
 \non
 &+& \Big(\delta \gamma_{ij}+2 \, \delta K_{\ha ij} \, x^\ha+  x^\ha x^{\hc}\big(\R_{i\ha \hc j}|_\Sigma +{2\over R} \, \delta^{\hat 2}_\hc  \, \delta K_{\ha\,i j} 
 - \delta^{\hat 2}_\ha \delta^{\hat 2}_\hc{\delta \gamma_{ij} \over R^2}  \big) \Big)dy^idy^j + \mathcal{O}(\epsilon^2)~ .
 \non
 \labell{spert}
\eea
Here $y^i$ are just the standard spherical angles multiplied by $R$. In what follows we use the unperturbed induced metric $\gamma_{ij}$ to raise and lower the indices on the entangling surface.

To use \reef{varEE} we need the connected correlator $\langle \hat T^{\mu\nu}_\mathcal{H}\hat H \rangle_c$. Since the Hamiltonian is conserved and hyperbolic space is maximally symmetric, the correlator is insensitive to where the operators are inserted, and therefore it is constant on $\mathcal{H}$. In particular, it was shown in \cite{eric} that
 \be
 \langle \hat T_{\mathcal{H} }^{\tau_\mt{E} \tau_\mt{E}} \hat H\rangle_c=-{(d-1)\over 2^{d+2}\pi^2 d}\, {\Omega_{d+2}\over R^{d+1}}\,C_T~,
 \quad \Omega_{d}={2\pi^{d+1\over 2}\over \Gamma\big({d+1\over 2}\big) }~,
 \labell{emtcorr1}
 \ee
where $C_T$ is a ``central charge'' common to CFTÕs in any number of dimensions. In four dimensions this coefficient is related to the standard central charge $c$ which appears as the coefficient of the (Weyl)$^2$ term in the trace anomaly\footnote{See  \reef{trace4}  for the definition of the central charges that we use throughout this paper.} $C_T = (40/\pi^4) c$.

Since the background geometry is conformally flat, all Weyl invariants of the trace anomaly vanish. Further, the background is the direct product of two lower dimensional geometries which dictates that the Euler density is also zero. Hence, the trace anomaly vanishes in this particular background. Using the tracelessness of the energy-momentum tensor and maximal symmetry of $H^{d-1}$ yields
 \be
 \langle \hat T_{\mathcal{H}  \, j}^{~i} \hat H\rangle_c={\delta_{\,j }^{i}\over 2^{d+2}\pi^2 d}\, {\Omega_{d+2}\over R^{d+1}}\,C_T~,
  \labell{emtcorr2}
 \ee
where indices $i,j$ run over the hyperbolic space $H^{d-1}$. It follows from   \reef{varEE} that the off diagonal elements of  \reef{spert} do not contribute to linear order corrections since the connected correlator $\langle \hat T^{\mu\nu}_\mathcal{H}\hat H \rangle_c$ is diagonal. 

Eqs. \reef{varEE}, \reef{spert}, \reef{emtcorr1} and \reef{emtcorr2} give a general solution for linear perturbations of spherical regions in flat space. In the next subsection we carry out a particular calculation in $d=4$ and show that our formula \reef{varEE} agrees with the known results in the literature.

\subsection{Calculation} \label{sec:calc}

Let us evaluate the logarithmic divergence of entanglement entropy for a four dimensional CFT using our result \reef{varEE}. This divergence is universal since it is independent of the details of regularization scheme, and it was shown in \cite{CHM} that for a perfect sphere in flat space it is entirely fixed by the coefficient of the A-type trace anomaly. In particular, in $d=4$ the universal divergence takes the form
\be
S_{univ}=-4a \log (R/\delta)~,
\labell{EEsph}
\ee
Here $\delta$ is the UV cut-off and $a$ is the central charge defined in \reef{trace4},

As argued in \cite{CHM}, the leading order term in   \reef{varEE} satisfies $S_T=S_{univ}$.  The logarithmic divergence within the thermal computation on $\mathcal{H}$ is a result of the divergent volume of hyperbolic space. This IR divergence emerges because we have a uniform entropy density, but the volume of $H^{3}$ is infinite. Hence, to regulate the thermal entropy in $\mathcal{H}$ we integrate to some maximum radius, $u = u_{max}$ where $u_{max} \gg 1$. On the other hand, the divergence of entanglement entropy is entirely due to short distance fluctuations in the vicinity of $\Sigma$. Thus, in order to regulate this divergence we exclude the $\delta$-neighborhood of the entangling surface $\Sigma$, where $\delta/R\ll 1$. These two UV and IR cut-off's should be related by the conformal mapping between the two spaces. If we focus on the $t _{\mt E}= 0$ slice (or equivalently the $\tau_{\mt E} = 0$ slice), then \reef{magic2}, yields the following relation \cite{CHM}
\be
 1-{\delta\over R}= \frac{\sinh u_{max}}{\cosh u_{max} + 1}~ \Rightarrow ~ u_{max}\simeq\log (R/\delta) . 
\ee
To get corrections to the leading order result we substitute eqs. \reef{spert}, \reef{emtcorr1} and \reef{emtcorr2} into \reef{varEE} and use eqs. \reef{hyper2}, \reef{confact} and \reef{magic2} to carry out the integrals. The final answer for the logarithmically divergent part of the integrals is given by
\be
 \delta S_{univ}={c\over 6\pi}\int_{\Sigma} \({\delta\gamma\over R^2}-{2\over R}  \, \delta^{\hat 2}_\hc \, \delta K^\hc 
 + \gamma^{ij}\delta^{\ha\hc} \R_{i\hat \ha\hc j}|_\Sigma + \delta^{ac}\delta^{bd}\R_{abcd} \) \log(R/\delta)~,
 \labell{4dcorr}
\ee
where $\Sigma$ is a sphere of radius $R$,  $\delta\gamma$ and $\delta K^\hc$ are the traces of the perturbations $\delta\gamma_{ij}$ and  $\delta K^\hc_{ij}$, and we used   \reef{magic2} to evaluate the components of $h_{\mu\nu}$ in coordinates \reef{hyper2},
\bea
 h_{uu}&=&-{R^2\, \Omega^4\over 6}  \delta^{ac}\delta^{bd}\R_{abcd} \, e^{-2 u} \sin^2 (\tau_{\mt E}/R)~,
 \non
 h_{\tau_{\mt E}\tau_{\mt E}}&=&-{R^2 \, \Omega^4\over 6} \delta^{ac}\delta^{bd}\R_{abcd}  \, \big(1+ e^{-u} \cos (\tau_{\mt E}/R) \big)^2~,
 \non
 h_{ u \tau_{\mt E} }&=&{R^2 \, \Omega^4 \over 24} \delta^{ac}\delta^{bd}\R_{abcd} ~ e^{- u} \sin (\tau_{\mt E}/R)\big(1+ e^{-u} \cos (\tau_{\mt E}/R) \big)~.
\eea

Let us now compare \reef{4dcorr} with Solodukhin's formula \reef{4dEE}. For the case of a sphere in flat space, this formula reduces to  \reef{EEsph}. Corrections  to  \reef{EEsph} can be evaluated by varying \reef{4dEE} around sphere of radius $R$ embedded into $R^d$. Provided that variations are small and satisfy  \reef{varparam}, we get \reef{4dcorr2} again. The latter is not a coincidence, it is a straightforward consequence of the fact that \reef{4dEE} is Weyl invariant while the two setups (a plane and a sphere in flat space) are conformally equivalent. To see it explicitly, let us write the metric around flat plane as follows
\be
 ds^2=dr^2+r^2d\theta^2 + \delta_{ij}dy^idy^j={ r^2 \over R^2}\Big(d\tau_{\mt E}^2 + {R^2\over r^2}(dr^2+ \delta_{ij}dy^idy^j)\Big)~,
\ee
where we have defined $\tau_{\mt E}=R\,\theta$ and used polar coordinates in the transverse space to the plane. Stripping off conformal factor on the right hand side of this expression leaves us with the metric on $\mathcal{H}$ in Poincare patch. Note that conformal factor is everywhere regular in the punctured Euclidean space (or analytically continued Rindler wedge), and the plane at $r=0$ is mapped onto conformal boundary of $\mathcal{H}$.

Hence, we have shown that two setups are conformally equivalent to $\mathcal{H}$ with entangling surfaces being mapped onto conformal boundary of the hyperbolic space. Therefore they are conformally equivalent to each other. In particular, it follows that quadratic in extrinsic curvatures term of \reef{4dEE},
\be
 I=\int_\Sigma (K^\ha_{ij} K^{ ij}_\ha-{1\over 2}K^\ha K_{\ha})~,
\ee
does not contribute to the first variation of entanglement entropy around spherical region. This claim can be verified by direct computation, however there is a simple argument based on the Weyl symmetry inherent to the problem. Indeed, this term is separately Weyl invariant and its first variation vanishes in the planar case, therefore the same is true for conformally equivalent  spherical region in flat space. In our forthcoming publication  we are going to explore the second order perturbation theory to uncover the effect of extrinsic curvatures on the entanglement entropy.

Let us now show that \reef{4dcorr2} agrees with \reef{4dcorr}. Varying the Gauss-Codazzi relation \reef{gc} around the unperturbed sphere of radius $R$ embedded in flat space gives
\be
 \gamma^{ij}\gamma^{kl} \R_{ikjl}|_\Sigma={1\over R^2}\delta\gamma-{2\over R} \, \delta_\hc^{\hat 2} \, \delta K^\hc 
 + \nabla^i(\nabla^j\delta\gamma_{ij} - \gamma^{mn}\nabla_i \, \delta\gamma_{mn})~,
 \labell{Rsig}
\ee 
where we have used the  variational rule \reef{varule}. Substituting this result into   \reef{4dcorr2} gives \reef{4dcorr}\footnote{The total derivative in   \reef{Rsig} does not contribute since $\Sigma$ in our case has no boundaries.}.

\acknowledgments
We thank Raphael Bousso, Ben Freivogel, Markus Luty, and Rob Myers for helpful discussions. This work is supported by the Berkeley Center for Theoretical Physics.

\appendix

\section{Notation}
\labell{not}

In this appendix we explain our notation and conventions. Greek indices run over the entire  background, whereas Latin letters from the `second' half of the alphabet $i,j,\ldots$ represent directions along the entangling surface.

There is a pair of independent orthonormal vectors which are orthogonal to $\Sigma$, we denote them by $n_{\ha}^\mu$ (with $\ha=1,2$), where the letters from the beginning of the Latin alphabet are used to denote the frame or tangent indices in the transverse space. Then delta Kronecker $\delta_{\ha \hb}=n_{\ha}^\mu n_{\hb}^{\nu}g_{\mu\nu}$ is the metric in the tangent space spanned by these vectors and $\delta^{\ha \hb}$ is the inverse of this metric. 

We also have tangent vectors
$t_i^\mu$ to $\Sigma$, which are defined in the usual way with
$t_i^\mu=\partial x^\mu/\partial y^i$, where $x^\mu$ and
$y^i$ are the coordinates in the full embedding space and along
the surface, respectively. The induced metric is then given by
$\gamma_{ij}=t^\mu_i\, t^\nu_j\, g_{\mu\nu}$. It can  also be defined as a bulk tensor with
$\gamma_{\mu\nu}=g_{\mu\nu}-g^{\perp}_{\mu\nu}$, where $g^{\perp}_{\mu\nu}=\delta_{\ha\hb}n^{\ha}_\mu n^{\hb}_\nu$ is the metric in the space transverse to $\Sigma$. The second
fundamental forms are defined for the entangling surface with $K^{\ha}_{i j} = t^\mu_i\, t^\nu_j \nabla_\mu n^{\ha}_{\nu}$, where $\nabla_\mu$ is covariant derivative compatible with $g_{\mu\nu}$. We use this definition to construct the bulk vector $K^{\mu}_{i j}=n^\mu_\ha K^{\ha}_{i j}$. 

Next we define the volume form in the tangent space spanned by the normal vectors
\bea
 \veps_{\ha\hb}&=&-\veps_{\hb\ha}~,\quad \veps_{\hat 1\hat 2}=1~,
 \non
 \veps^{\ha\hb}&=&\delta ^{\ha\hc}\delta^{\hb\hd}\veps_{\hc\hd}= \veps_{\ha\hb}~.
 \labell{vol}
\eea
Using this definition the volume form in the transverse space can be written as $\veps_{\mu\nu}=\veps_{\ha\hb} n^{\ha}_\mu n^{\hb}_\nu$. We use $g^\perp_{\mu\nu}$ to raise and lower the indices in the transverse space, while indices along the direction of the entangling surface are raised and lowered with the induced metric $\gamma_{\mu\nu}$. Note that the following useful identity holds,
\be
 \veps_{\mu\nu}\veps_{\rho\sigma}=g^{\perp}_{\mu\rho}g^{\perp}_{\nu\sigma} - g^{\perp}_{\mu\sigma}g^{\perp}_{\nu\rho}~.
 \labell{Vform}
\ee
Finally, our convention for the curvature tensor is given by
\be
 \R_{\mu\nu\rho\sigma}={1\over 2}(g_{\mu\sigma,\nu\rho}+g_{\nu\rho,\mu\sigma}-g_{\mu\rho,\nu\sigma}-g_{\nu\sigma,\mu\rho})
 +\Gamma_{\nu\rho,\chi}\Gamma^\chi_{\mu\sigma}-\Gamma_{\nu\sigma,\chi}\Gamma^{\chi}_{\mu\rho}~.
 \labell{defRiem}
\ee

\section{Foliation of $\M$ in the vicinity of the entangling surface}
\labell{fol}

In this appendix we build a particular foliation of $\cM$ in the vicinity of $\Sigma$. First, we choose some parametrization $\{y^i\}_{i=1}^{d-2}$ for the entangling surface $\Sigma$, then for a given point $O\in\Sigma$ we fill the transverse space with geodesics radiating orthogonally out from $O$. For each point $p$ on the resulting two-dimensional manifold, $T_O$, we find a geodesic that connects it to $O$, such that $p$ lies a unit affine parameter from $O$. Tangent vector to such a geodesic at $O$ can be expanded in terms of a chosen two-dimensional basis $n_{\ha}^\mu$. We give its components the names $x^{\ha}$ and choose them as coordinates on $T_O$. Together $\{y^i,x^\ha\}$ parametrize $\cM$ in the vicinity of $\Sigma$. 

Note that we keep the parametrization of the entangling surface unspecified and therefore the final answer for entanglement entropy will be symmetric with respect to reparametrizations of $\Sigma$. On the other hand, choosing a particular foliation of the transverse space does not destroy general covariance of the entanglement entropy since  the final answer is obtained by integrating out this space.  

By construction, the following relations hold  
\be
 n^\ha_{\mu}=\delta_{\mu}^\ha ~, \quad t^\mu_i=\delta_i^\mu~, \quad g^{\perp}_{\mu\nu}=\delta_{\ha\hc} \, \delta_{\mu}^\ha \, \delta_{\nu}^\hc~,\quad g_{i\ha}=0 \text{~on~} \Sigma~.
 \labell{normal}
\ee
In particular, $\delta_{\ha\hc}$ plays the role of the transverse metric in this foliation and one can readily evaluate the extrinsic curvatures of $\Sigma$,
\be
 K_{ij}^\ha= \nabla_i n^\ha_j\big|_\Sigma={1\over 2}\delta^{ \ha\hc}\del_{\hc} \, g_{ij}\big|_\Sigma~.
\ee
Hence, 
\be
g_{ij}=\gamma_{ij}+2 K_{\ha ij} \, x^\ha+\mathcal{O}(x^2)~.
\labell{sigmet}
\ee 
Furthermore, geodesics radiating orthogonally out from a given point $y\in\Sigma$ take the form $x^\ha(\tau)=v^\ha \tau$, where $v^\ha$ belongs to the two-dimensional tangent space spanned by two normal vectors at $y$. Substituting this parametrization into the geodesic equation yields
\be
\Gamma^\mu_{\ha\hc}v^\ha v^\hc=0 ~ \Rightarrow ~ \Gamma^\mu_{\ha\hc}=0 ~~ \text{at} ~~  O ~.
\ee
This identity can be further generalized by differentiating the geodesic equation $n$ times with respect to $\tau$ and setting $\tau=0$. This gives
\be
 \del_{( \hd_1}\del_{\hd_2}\cdots\del_{\hd_n} \Gamma^\mu_{\ha\hc)}=0 ~~ \text{at} ~~  O ~,
 \labell{eq:B5}
\ee 
where as usual $(\cdots)$ denotes symmetrization with respect to the indices within the parenthesis. This result (\ref{eq:B5}) with index $\mu$ in the transverse space can be used to derive the expansion of the metric on $T_y$,
\bea
 g_{\ha\hb}(x,y)&=&\delta_{\ha\hb}-{1\over 3}\R_{\ha\hc\hb\hd}(y)x^{\hc}x^{\hd}-{1\over 6}\del_\he \R_{\ha\hc\hb\hd}(y)\,x^{\hc}x^{\hd}x^\he+ \mathcal{O}(x^4)~.
  \labell{transmet}
\eea
Moreover, it follows from $\Gamma^i_{\ha\hc}|_\Sigma=\del_{(\hd}\Gamma^i_{\ha\hc)}|_\Sigma=0$ that Taylor expansion of $g_{i\hc}$ in the vicinity of $\Sigma$ can be written as follows
\be
  g_{i\hc}= \big(A_i+{1\over 3}x^\hb \veps^{\hd\he}\R_{i\hb\hd\he}\big|_{\Sigma}  \big) \, x^\ha  \veps_{\ha\hc}    + \mathcal{O}(x^3)~,
 \labell{offdiag}
\ee
where we have introduced a vector field that lives on $\Sigma$
\be
 A_i={1\over 2}\veps^{\ha\hc} \del_\ha g_{i \hc}\big|_{\Sigma}~,
\ee
and used the following identity that holds for our foliation
\be
 \R_{i\hb\ha\hc}\big|_{\Sigma}=\del_{\hb}\del_{[\ha}\, g_{\hc] i}\big|_{\Sigma}~,
\ee
where $[\cdots]$ denotes antisymmetrization with respect to the indices inside the square brackets.

We only need to compute $\mathcal{O}(x^2)$ term in  \reef{sigmet} to get the expansion of the full metric to second order in the distance from the entangling surface. We first note that Christoffel symbols with at least one index in the transverse space are given by
\be
\Gamma^\mu_{\ha\hc}|_\Sigma=0 ~,\quad \Gamma^\ha_{i\hc}|_\Sigma=-\veps^\ha_{~\hc} A_i ~, 
\quad \Gamma_{ij}^\ha|_\Sigma=-K_{ij}^\ha~, \quad \Gamma_{i\ha}^j|_\Sigma=K_{\ha\,i}^{j}~.
 \labell{Christoff}
\ee
Now using  \reef{defRiem}, we obtain
\be
\R_{i\ha j\hb}|_\Sigma=  {1\over 2} \veps_{\hb\ha} F_{ij}-{1\over 2}\del_\ha\del_\hb g_{ij}|_\Sigma+\delta_{\ha\hb} A_iA_j
+K_{\hb \, i l} K_{\ha\,j}^{~l}~,
\labell{mixriem}
\ee
where $F_{ij}=\del_iA_j-\del_jA_i$ is the field strength. Symmetrizing this expression with respect to $\ha$ and $\hb$, yields
\be
 {1\over 2} \del_\ha\del_\hb g_{ij}|_\Sigma= \R_{i(\ha \hb) j}|_\Sigma+\delta_{\ha\hb} A_iA_j+{1\over2}(K_{\hb \, i l} K_{\ha\,j}^{~l}+K_{\ha \, i l} K_{\hb\,j}^{~l})~,
\ee
where $(\cdots)$ means symmetrization with respect to the indices inside the parenthesis. Hence  \reef{sigmet} to second order in $x^\ha$ takes the form
\be
g_{ij}=\gamma_{ij}+2 K_{\ha ij} \, x^\ha+x^\ha x^{\hc}\big( \delta_{\ha\hc} A_iA_j+\R_{i(\ha \hc) j}|_\Sigma \big)+x^\ha x^{\hc}K_{\hc \, i l} K_{\ha\,j}^{~l}+\mathcal{O}(x^3)~.
\labell{sigmet2}
\ee 
Altogether eqs. \reef{transmet}, \reef{offdiag} and \reef{sigmet2} correspond to the second order expansion of the full metric $g_{\mu\nu}$ in the vicinity of $\Sigma$. To linear order in the distance from $\Sigma$ this metric takes the simple form, 
\be
ds^2=\delta_{\ha\hc}dx^\ha dx^\hc+2A_i \veps_{\ha\hc} \, x^\ha dx^\hc \, dy^i +(\gamma_{ij}+2 K_{\ha ij} \, x^\ha)dy^idy^j
+ \mathcal{O}(x^2)~.
  \labell{ans0}
\ee

Note that using the definition \reef{defRiem} and   \reef{Christoff}, one can evaluate various components of the Riemann tensor that were not necessary so far. For instance, considering directions along the entangling surface $\Sigma$ yields the well known Gauss-Codazzi identity
\be
\R_{ijkl}|_\Sigma=\R_{ijkl}^{\Sigma}+K_{jk}^\ha K_{\ha\,il}-K_{jl}^\ha K_{\ha\,ik}~,
\labell{gc}
\ee
where $\R_{ijkl}^{\Sigma}$ is the intrinsic curvature tensor on $\Sigma$. 

Furthermore,
\bea
\R_{ij\ha\hb}|_\Sigma&=&\veps_{\hb\ha} F_{ij}
+K_{\hb \, i l} K_{\ha\,j}^{l} - K_{\ha \, i l} K_{\hb\,j}^{l}~,
\labell{mixriem0}
\eea
This identity can be used to express the field strength in terms of the background curvature and extrinsic geometry of $\Sigma$. 

Finally,
\bea
 \R_{ijl\ha}|_\Sigma&=&  \nabla_{i} K_{\ha j l}-\nabla_{j} K_{\ha i l}+2\,\veps_{\hb\ha}A_{[i} K^\hb_{j]l}~,
 \non
 \R_{\ha\hb\hc\hd}|_\Sigma&=&\R^T_{\ha\hb\hc\hd}|_\Sigma~,
\eea
where $\nabla_i$ is the covariant derivative compatible with the induced metric on $\Sigma$ and $R^T_{\ha\hb\hc\hd}$ is the intrinsic curvature tensor of the transverse space, $T_y$, at a given point $y\in\Sigma$.

\section{Intermediate calculations for Sec.~\ref{rindler} }
\label{sec:appC}

In this Appendix we evaluate the integral (\ref{eq:deltaS2}) appearing in the calculation of the first order correction to the entanglement entropy for a deformed plane in a weakly curved background. First we consider the contribution of the metric perturbation with indices in the direction of the entangling surface, \ie $h_{ij} = x^{a}x^{c}R_{i a c j}$. In this case (\ref{eq:OsI}) is given by,
\begin{equation}
\mathcal{I}_{ij,22} = \frac{4\ x_2^2\ (y-\bar{y})_i (y-\bar{y})_j}{\left((x- \bar{x})^2 +(y-\bar{y})^2\right)^2} - \frac{1}{4} \delta_{i j}
\end{equation}
We begin  evaluating (\ref{eq:deltaS2}) by first doing the integral over $\bar{y}$ through a change of variables $\bar{y} \rightarrow \bar{y}+y$ giving
\begin{equation} \label{eq:deltaS3}
\delta S_1 = - {\pi^2\over 10} \, C_T \int_{\bar{x}_1>0}{d^2x \, d^2y \, d\bar{x}_1\,  \bar{x}_1 \, 
 \frac{ \delta^{ij}\, h_{i j} }{ ((x_1-\bar{x}_1)^2 +x_2^2)^3}\left(\frac{ x_2^2}{(x_1-\bar{x}_1)^2 +x_2^2}- \frac{5}{6}\right)} .
\end{equation}
Next, we carry out the $\bar{x}_1$ integral and introduce polar coordinates in the transverse space, $x_1 = r \cos \theta$, $x_2 = r\sin\theta$, 
\be
\delta S_1 =  {\pi^2\over 240} \, C_T \int{d^2 y \, d\theta \, { dr\over r^3}\,    \delta^{ij} \, h_{ij }},
\label{eq:deltaS4}
\ee
As expected, the integral over $r$ exhibits logarithmic divergence close to the entangling surface at $r=0$. Hence, we introduce a UV cut off, $\delta$, to regularize divergence and integrate over $r$ and $\theta$
\begin{equation} \label{eq:deltaS1}
\delta S_1 = \frac{c}{6 \pi}   \int{d^2y\,\, \delta^{ij} \delta^{ac}\, \mathcal{R}_{i a c j}  } \log(\ell/\delta)~,
\end{equation}
where $\ell$ is characteristic scale of small perturbations, and we used the value of $C_T=(40/\pi^4) c$ in four spacetime dimensions.

Next we calculate the contribution of perturbations in the transverse space, \ie $h_{ab} = -\frac{1}{3}R_{a c b d}x^{c}x^{d}$. Using $\mathcal{I}_{ab,22}$  from (\ref{eq:OsI}) 
and performing the integral over $\bar{y}$ in (\ref{eq:deltaS2}) yields
\begin{eqnarray}
\delta S_2 &=& -\pi^2 C_T \int_{\bar{x}_1>0}{d^2x\ d^2y\ d\bar{x}_1\  \bar{x}_1\ h_{a b}(x,y)}  \nonumber \\
&\times& \left(\frac{1}{3}\frac{\delta_{a 2}\delta_{b 2} - \delta_{ab}/4 }{\left((x_1-\bar{x}_1)^2 +x_2^2\right)^3} - \frac{x_2 (x - \bar{x})_b \, \delta_{a 2} }{\left((x_1-\bar{x}_1)^2 +x_2^2\right)^4}+ \frac{4 }{5} \, \frac{x_2^2\, (x-\bar{x})_a (x-\bar{x})_b}{\left((x_1-\bar{x}_1)^2 +x_2^2\right)^5}\right) \nonumber
\end{eqnarray}
As before, we preform the $\bar{x}_1$ integral, introduce polar coordinates in the transverse space, substitute $h_{ab}$, carry out $\theta$ integral, and finally get
\begin{equation}
\delta S_2  = \frac{c}{6 \pi} \int{d^2 y\ \delta^{ac}\delta^{bd}\R_{abcd}}  \log(\ell/\delta) 
\end{equation}
Combined with (\ref{eq:deltaS1}), we have thus recovered (\ref{eq:RindAnswer}).

\end{document}